\documentclass[twocolumn]{webofc}
\usepackage[varg]{txfonts}   
\usepackage{bm}
\usepackage{mathtools}
\usepackage{textcomp}
\newcommand{\Ri}{R_\mathrm{i}}
\newcommand{\Ro}{R_\mathrm{o}}
\newcommand{\rhoi}{\rho_\mathrm{i}}
\newcommand{\sinc}{\mathrm{sinc}}

\begin{document}
\title{Stress distribution in elastic disks with a hole under uniaxial compression}
\author{
    \firstname{Ken} \lastname{Okamura}\inst{1}\fnsep
    \thanks{\email{okamurak@st.go.tuat.ac.jp}} \and
    \firstname{Yosuke} \lastname{Sato}\inst{2} \and
    \firstname{Satoshi} \lastname{Takada}\inst{2}\fnsep
    \thanks{\email{takada@go.tuat.ac.jp}, corresponding author}
}

\institute{
    Department of Industrial Technology and Innovation, 
    Tokyo University of Agriculture and Technology, 
    2-24-16 Naka-cho, Koganei, Tokyo 184-8588, Japan
    \and
    Department of Mechanical Systems Engineering, 
    Tokyo University of Agriculture and Technology, 
    2-24-16 Naka-cho, Koganei, Tokyo 184-8588, Japan
}

\abstract{
This paper investigates the stress and displacement distribution in a two-dimensional elastic hollow disk subjected to distributed diametric loading, extending our previous analysis of concentrated loading [Okamura et al. \textit{Strength Mater.}~\textbf{57}, 102--114 (2025)]. 
The study provides deeper insights into the mechanical behavior of materials such as concrete and rock by examining the effects of load distribution on stress localization and displacement patterns. 
Using elastodynamic theory, we derive the static stress distributions and identify key differences from the concentrated loading case, particularly in the locations and magnitudes of stress extrema. 
This work contributes to a more comprehensive understanding of stress behavior in elastic disks under realistic loading conditions.
}
\maketitle


\section{Introduction}
The mechanical strength of concrete and rock materials plays a crucial role in the design and construction of durable structures. 
Among various methods used to assess tensile strength, the Brazilian test remains one of the most widely employed techniques \cite{Hobbs, Claesson22}. 
This method involves applying a diametric load to a cylindrical specimen, generating tensile stress perpendicular to the loading axis. 
Despite its effectiveness, the test is affected by significant shear stress concentrations near the points of load application, potentially influencing the accuracy of the measured tensile strength \cite{Timoshenko, Jaeger, Ramesh22}.
Usually, concrete and rock behave inelastically, making it difficult to apply the theory of elasticity. 
However, in the case of quasi-static experiments, the elasticity can be applied because materials behave elastically from the beginning of deformation until failure~\cite{Hobbs, Timoshenko}.

To mitigate these stress concentrations, the ring test has been proposed, where a central hole is introduced in the specimen \cite{Hondros, Laurent, Gao}. While previous research has provided insights into stress distribution along the loading axis, a complete understanding of the overall stress field within the disk is still lacking. Investigations into two-dimensional elastic hollow disks have presented stress distributions \cite{Ujihashi74, Hiramatsu70}, yet detailed theoretical derivations and validation of these results remain limited.

Analytical approaches to stress distribution in such systems often rely on stress functions and complex variable techniques, which are well-suited for static, two-dimensional elasticity problems \cite{Muskhelishvili, Timoshenko, Schonert04, Wang85, Wu06}. 
However, incorporating elastodynamic theory allows for the consideration of temporal stress and displacement variations, distinguishing between different wave effects. This approach has been successfully applied to elastic disks and spheres, providing insights into dynamic stress responses \cite{Love, Eringen, Jingu85, Ujihashi74, Kessler91, Hua19}.

In this work, we revisit the stress analysis of two-dimensional elastic hollow disks through the lens of elastodynamic theory. By examining the transition from dynamic to static conditions, we aim to provide a more comprehensive characterization of stress distributions under diametric loading.

\section{Model and setup}
\begin{figure}[htbp]
    \centering
    \includegraphics[width=\linewidth]{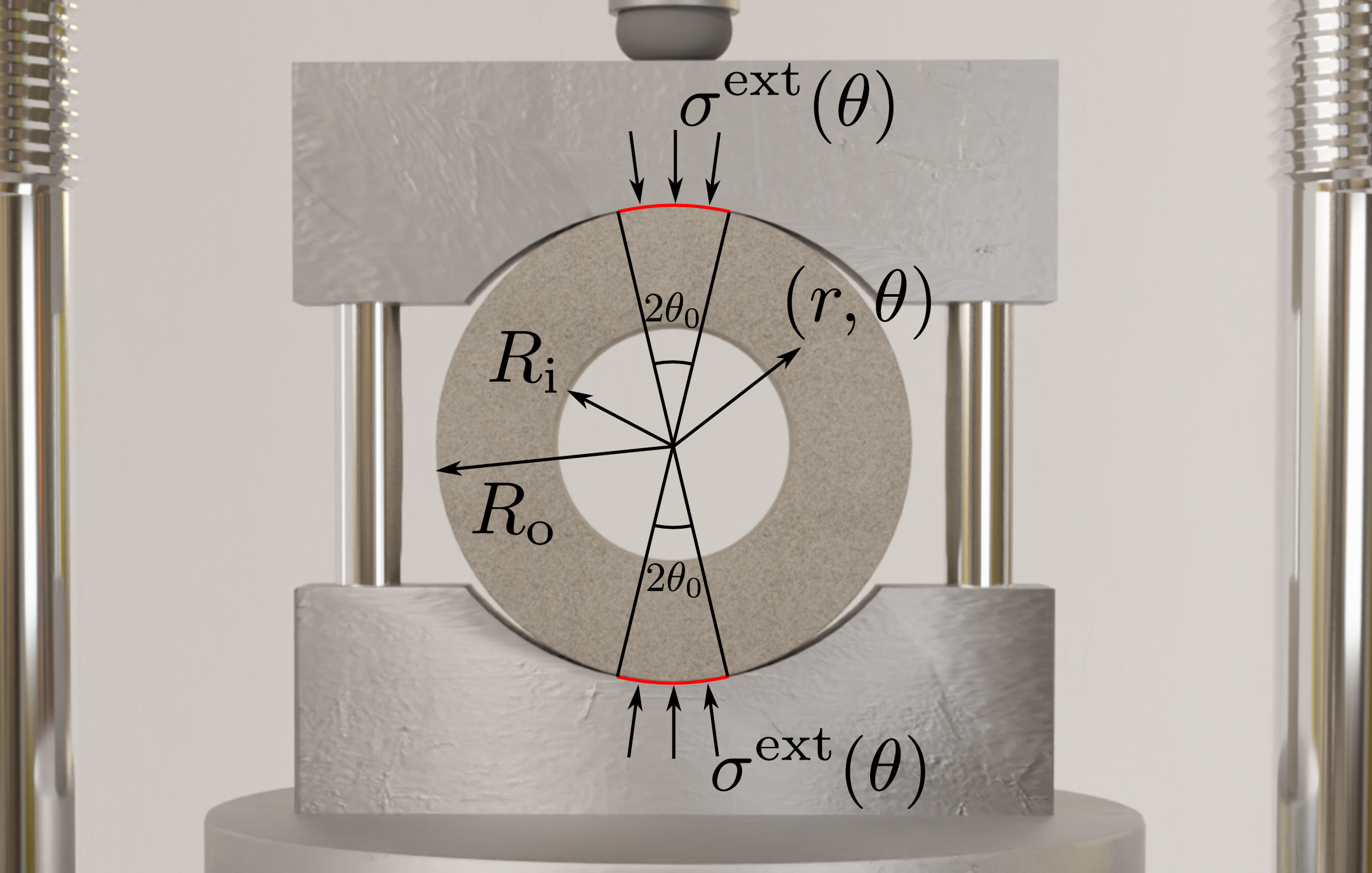}
    \caption{Schematic of our system.
    The stress acts on the outer surface of a hollow disk whose outer and inner radii are given by $\Ro$ and $\Ri$, respectively.}
    \label{fig:setup}
\end{figure}

Consider a two-dimensional elastic hollow disk with outer and inner radii denoted as $\Ro$ and $\Ri$, respectively. 
At time $t=0$, a diametric load is applied to the disk, as illustrated in Fig.~\ref{fig:setup}.
Without loss of generality, we define these two loading points at $\theta=0$ and $\pi$. 
Under these conditions, we analyze the deformation and stress within the disk. 
The disk must satisfy the following boundary conditions:
\begin{subequations}\label{eq:boundary_condition}
\begin{align}
    \left.\sigma_{rr}\right|_{r=R_\mathrm{o}}
    &=-\sigma^\mathrm{ext}(\theta)\Theta(t),\\
    \left.\sigma_{r\theta}\right|_{r=R_\mathrm{o}}
    &= \left.\sigma_{rr}\right|_{r=R_\mathrm{i}}
    = \left.\sigma_{r\theta}\right|_{r=R_\mathrm{i}}=0,
\end{align}
\end{subequations}
where $\sigma^\mathrm{ext}(\theta)$ can be expanded as a Fourier series, assuming it has a period of $\pi$ with respect to $\theta$, and can be written as
\begin{equation}
    \sigma^\mathrm{ext}(\theta)
    = 2\sigma_0
    \left[\frac{\widetilde{\varsigma}_0}{2}
    + \sum_{m=2,4,\cdots}\widetilde{\varsigma}_m \cos(m\theta)\right].
\end{equation}
Here, the coefficients $\widetilde{\varsigma}_m$ will be considered in detail later.

\section{Linear elastodynamics}\label{sec:elastodynamics}
Let us start our analysis from a linear elastodynamic equation.
Assuming the plane stress condition, the deformation vector $\bm{u}$ satisfies the Navier-Cauchy equations \cite{Fung}:
\begin{equation}
    \varrho_0\frac{\partial^2}{\partial t^2}\bm{u}
    = G\nabla^2 \bm{u} + G\frac{1+\nu}{1-\nu}\bm\nabla \left(\bm\nabla \cdot \bm{u}\right),
    \label{eq:Navier-Cauchy}
\end{equation}
where $G$ is the shear modulus, $\nu$ is Poisson's ratio, and $\varrho_0$ is the mass density.
Next, we introduce the longitudinal and transverse velocities as$v_\mathrm{L}\equiv [2G/[(1-\nu)\varrho_0]^{1/2}$ and $v_\mathrm{T}\equiv (G/\varrho_0)^{1/2}$, respectively, and define the following dimensionless quantities for convenience:
\begin{align}
    &\bm{\rho} \equiv \frac{\bm{r}}{\Ro},\quad
    \rhoi \equiv \frac{\Ri}{\Ro},\quad
    \tau \equiv \frac{v_{\rm L}t}{\Ro},\quad
    \widetilde{\bm{u}} \equiv \frac{G}{\sigma_0}\frac{\bm{u}}{\Ro},\nonumber\\
    &\overleftrightarrow{\widetilde{\sigma}} 
    \equiv \frac{\overleftrightarrow{\sigma}}{\sigma_0},\quad
    \mu\equiv \frac{v_\mathrm{L}}{v_\mathrm{T}}
    =\sqrt{\frac{2}{1-\nu}}(>1).
\end{align}
Thus, the Navier-Cauchy equation~\eqref{eq:Navier-Cauchy} can be rewritten in dimensionless form as:
\begin{equation}
    \mu^2\frac{\partial^2}{\partial \tau^2}\widetilde{\bm{u}}
    = \nabla_{\bm{\rho}}^2 \widetilde{\bm{u}} 
    + \frac{1+\nu}{1-\nu}\bm\nabla_{\bm{\rho}} \left(\bm\nabla_{\bm{\rho}} \cdot \widetilde{\bm{u}}\right),
    \label{eq:Navier-Cauchy2}
\end{equation}
where $\bm\nabla_{\bm{\rho}}\equiv R_\mathrm{o} \bm\nabla$.

From Helmholtz's theorem, the dimensionless deformation vector can be written in terms of the scalar and vector potentials, $\phi$ and $\bm{A}=(0,0,A)^\mathrm{T}$, respectively, as
\begin{equation}
    \bm{\widetilde{u}} 
    =\bm\nabla_{\bm{\rho}} \phi 
    + \bm\nabla_{\bm{\rho}} \times \bm{A}.
\end{equation}
Substituting this into the Navier-Cauchy equation~\eqref{eq:Navier-Cauchy2}, we find that both potentials satisfy the wave equations \cite{Fung}:
\begin{equation}
    \nabla_{\bm{\rho}}^2 \phi 
    = \frac{\partial^2\phi}{\partial \tau^2},\quad
    \nabla_{\bm{\rho}}^2 A = \mu^2\frac{\partial^2A}{\partial \tau^2}.
    \label{eq:phi_A_wave_eq}
\end{equation}
To solve these wave equations, we introduce the Laplace transforms of $\phi$ and $A$ as
\begin{equation}
    \overline{\phi}(s)\equiv \int_0^\infty \phi\, \mathrm{e}^{-s\tau}\mathrm{d}\tau,\quad
    \overline{A}(s)\equiv \int_0^\infty A\, \mathrm{e}^{-s\tau}\mathrm{d}\tau.
    \label{eq:def_Laplace}
\end{equation}
Assuming that the system is static for $\tau<0$, the wave equations are solvable, and the general solutions are:
\begin{subequations}\label{eq:phi_bar_A_bar}
\begin{align}
    \overline{\phi}(s)
    &= \sum_{m=0}^\infty  
    \left[a_m^{I}(s) I_m (\rho s)
    + a_m^{K}(s) K_m (\rho s)\right] \cos(m\theta),\\
    \overline{A}(s)
    &= \sum_{m=1}^\infty  
    \left[b_m^I(s)I_m (\mu \rho s)
    + b_m^K(s)K_m (\mu \rho s)\right]\sin(m\theta),
\end{align}
\end{subequations}
where $I_m(r)$ and $K_m(r)$ are the modified Bessel functions of the first and second kinds, respectively~\cite{Abramowitz}.
The coefficients $a_m^I(s)$, $a_m^K(s)$, $b_m^I(s)$, and $b_m^K(s)$ should be determined to satisfy the boundary condition for the stress, given by Eq.~\eqref{eq:boundary_condition}.

For this purpose, let us first express the deformation vector and the stress tensor in terms of the potentials. 
By adopting the Laplace transforms of $\widetilde{\bm{u}}$ and $\overleftrightarrow{\widetilde{\sigma}}$ in the same manner as in Eq.~\eqref{eq:def_Laplace}, Similar infinite series forms for $\overline{u}$ and $\overleftrightarrow{\overline{\sigma}}$ can also be obtained, as in Eq.~\eqref{eq:phi_bar_A_bar} (see also Ref.~\cite{Okamura25} for the detailed derivation).

The stresses $\overline{\sigma}_{\rho\rho}$ and $\overline{\sigma}_{\rho\theta}$ should satisfy the boundary conditions \eqref{eq:boundary_condition} in the Laplace space.
Then, we can determine the coefficients as 
\begin{subequations}\label{eq:a_b}
\begin{align}
    a_m^I(s)&=
    \begin{cases}
        -\dfrac{1}{2s}\dfrac{G_{0,1}(\rhoi s)}{
        \renewcommand{\arraystretch}{0.5}
        \begin{array}{l}
            F_{0,1}(s)G_{0,1}(\rhoi s)\\
            \hspace{0.5em}-F_{0,1}(\rhoi s)G_{0,1}(s)
        \end{array}}
        \widetilde{\varsigma}_0& (m=0)\\
        0 & (m=1,3,\cdots)\\
        -\dfrac{N_{m,1}(s)}{sD_m(s)}\widetilde{\varsigma}_m & (m=2,4,\cdots)
    \end{cases},\\
    b_m^I(s)&=
    \begin{cases}
        0 & (m=1,3,\cdots)\\
        -\dfrac{N_{m,2}(s)}{sD_m(s)}\widetilde{\varsigma}_m & (m=2,4,\cdots)
    \end{cases},\\
    a_0^K(s) &=
    \begin{cases}
        \dfrac{1}{2s}\dfrac{F_{0,1}(\rhoi s)}{
        \renewcommand{\arraystretch}{0.5}
        \begin{array}{l}
            F_{0,1}(s)G_{0,1}(\rhoi s)\\
            \hspace{0.5em}-F_{0,1}(\rhoi s)G_{0,1}(s)
        \end{array}}
        \widetilde{\varsigma}_0 & (m=0)\\
        0 & (m=1,3,\cdots)\\
        -\dfrac{N_{m,3}(s)}{sD_m(s)}\widetilde{\varsigma}_m & (m=2,4,\cdots)
    \end{cases},\\
    b_m^K(s)&=
    \begin{cases}
        0 & (m=1,3,\cdots)\\
        -\dfrac{N_{m,4}(s)}{sD_m(s)}\widetilde{\varsigma}_m & (m=2,4,\cdots)
    \end{cases},
\end{align}
\end{subequations}
where $F_{m,i}(s)$, $G_{m,i}(s)$ ($i=0,\cdots,3$), $D_m(s)$, and $N_{m,i}$ that appear in the above equations are the same as those used in Ref.~\cite{Okamura25}.

Using Eqs.~\eqref{eq:a_b} and performing the inverse Laplace transforms, each component of the (dimensionless) deformation vector and the (dimensionless) stress tensor is expressed as
\begin{equation}
    \begin{Bmatrix}
        \widetilde{\bm{u}} \\
        \overleftrightarrow{\widetilde{\sigma}}
    \end{Bmatrix}
    = \frac{1}{2\pi \mathrm{i}}\int_\mathrm{Br}
    \begin{Bmatrix}
        \overline{\bm{u}} \\
        \overleftrightarrow{\overline{\sigma}}
    \end{Bmatrix}
    \mathrm{e}^{s\tau}\mathrm{d}s.
    \label{eq:u_sigma_inv_Laplace}
\end{equation}
It is important to note that $\int_\mathrm{Br}=\int_{\gamma-\mathrm{i}\infty}^{\gamma+\mathrm{i}\infty}$ represents the Bromwich integral used to compute the inverse Laplace transform. 
Here, $\gamma$ ($>0$) must be greater than the real parts of all poles of the integrands in Eq.~\eqref{eq:u_sigma_inv_Laplace}.

We focus on the static solutions of Eqs.~\eqref{eq:u_sigma_inv_Laplace}. 
These static solutions correspond to the long-time limit, $\tau\to\infty$, which can be evaluated using the final value theorem of Laplace transforms $\lim_{\tau\to\infty}f(\tau)=\lim_{s\to0}[s \overline{f}(s)]$.
After performing the necessary calculations, aided by l'Hopital's rule, we obtain the following static solutions:
\begin{subequations}\label{eq:u_sigma_0_static}
\begin{align}
    \widetilde{u}_\rho
    &= - \frac{(1-\nu)\rho^2+(1+\nu)\rhoi^2}{2(1+\nu)\rho(1-\rhoi^2)}\widetilde{\varsigma}_0\nonumber\\
    &\hspace{1em}
    + \sum_{m=2,4,\cdots} \frac{\mathcal{N}_m^{(1)}}{2(1+\nu)(m^2-1)\rho \mathcal{D}_m}\widetilde{\varsigma}_m\cos(m\theta),\\
    \widetilde{u}_\theta
    &= {\sum_m}^\prime \frac{\mathcal{N}_m^{(2)}}{2(1+\nu)(m^2-1)\rho \mathcal{D}_m}\widetilde{\varsigma}_m\sin(m\theta),\\
    \widetilde{\sigma}_{\rho\rho}
    &= -\frac{\rho^2-\rhoi^2}{\rho^2(1-\rhoi^2)}\widetilde{\varsigma}_0
    + \sum_{m=2,4,\cdots} \frac{\mathcal{N}_m^{(3)}}{\rho^2 \mathcal{D}_m}\widetilde{\varsigma}_m\cos(m\theta),\\
    \widetilde{\sigma}_{\rho\theta}
    &= \sum_{m=2,4,\cdots}
    \frac{\mathcal{N}_m^{(4)}}{\rho^2 \mathcal{D}_m}\widetilde{\varsigma}_m\sin(m\theta),\\
    \widetilde{\sigma}_{\theta\theta}
    &= -\frac{\rho^2+\rhoi^2}{\rho^2(1-\rhoi^2)}\widetilde{\varsigma}_0
    + \sum_{m=2,4,\cdots}
    \frac{\mathcal{N}_m^{(5)}}{\rho^2 \mathcal{D}_m}\widetilde{\varsigma}_m\cos(m\theta),
\end{align}
\end{subequations}
where $\mathcal{D}_m$ and $\mathcal{N}_m^{(i)}$ ($i=1,2,\cdots, 5$) are the same as those in Ref.~\cite{Okamura25}.
It should be noted that these solutions are consistent with the previous studies when consider the point force condition ($\varsigma_m=1/\pi$)~\cite{Okamura25}.

\section{Result}
In this section, we visualize the expressions for displacement and stress derived in the previous section. 
As the boundary condition applied externally, we consider a constant radial compression
\begin{equation}
    \sigma^\mathrm{ext}(\theta)=
    \begin{cases}
        \sigma_0 & (|\theta|\le \theta_0,\ |\theta-\pi|\le \theta_0)\\
        0 & (\text{otherwise})
    \end{cases},
\end{equation}
with the half width $\theta_0$ ($<\pi/2$).
In this case, the coefficient $\widetilde{\varsigma}_m$ becomes
\begin{equation}
    \widetilde{\varsigma}_m
    = 4\frac{\theta_0}{\pi}\sinc(m\theta_0),
\end{equation}
for $m=0, 2, 4, \cdots$, and $\widetilde{\varsigma}_m=0$ for $m=1, 3, 5, \cdots$, where $\sinc(x)\equiv \sin(x)/x$ is the sinc function.

\begin{figure}[htbp]
    \centering
    \includegraphics[width=\linewidth]{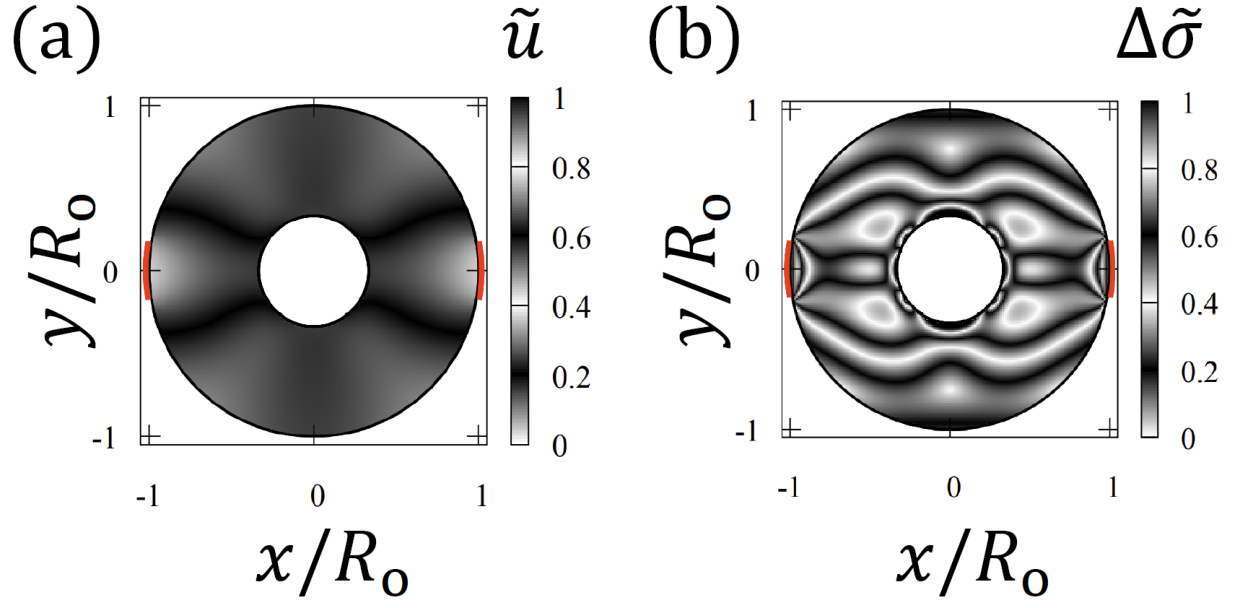}
    \caption{Plots of (a) the (dimensionless) displacement $\widetilde{u}$ and (b) the principal stress difference $\Delta \widetilde{\sigma}$ evaluated from Eq.~\eqref{eq:def_u_sigma} for $\nu=0.3$, $\rhoi=0.4$, and $\theta_0=\pi/18$.
    The thick lines at the outer edge represent the regions where the external stress stress acts.
    The color indicates the magnitude of each quantity.}
    \label{fig:static}
\end{figure}

Now, we introduce the magnitude of the dimensionless deformation $\widetilde{u}$ and the second stress difference $\Delta\widetilde{\sigma}$ as
\begin{equation}
    \widetilde{u}
    \equiv \sqrt{\widetilde{u}_\rho^2 + \widetilde{u}_\theta^2},\quad
    \Delta\widetilde{\sigma}
    \equiv \sqrt{\left(\widetilde{\sigma}_{\rho\rho}-\widetilde{\sigma}_{\theta\theta}\right)^2
    + 4\widetilde{\sigma}_{\rho\theta}^2},
    \label{eq:def_u_sigma}
\end{equation}
respectively.
The latter is known to be related to the interference fringes observed in photoelastic experiments~\cite{Coker57}.

The results of the (dimensionless) displacement $\widetilde{u}$ and the principal stress difference $\Delta \widetilde{\sigma}$ evaluated from Eq.~\eqref{eq:def_u_sigma} are shown in Fig.~\ref{fig:static} for $\nu=0.3$ and $\rhoi=0.4$ using a colormap by Python matplotlib.
The spatial distributions of both quantities are almost the same as those for the concentrated loading cases~\cite{Okamura25}.
We also note that these results are also consistent with the experimental result reported in Ref.~\cite{Tokovyy10}.
Although direct quantitative comparison is not feasible due to the absence of numerical data, our analytical results exhibit characteristic features that are consistent with the experimental observations. 
In particular, both the analytical and experimental results show a sharp variation in stress near the inner boundary of the disk, as well as horizontal stripe-like patterns extending across the interior region. 
These features indicate a strong similarity in the overall stress distribution and deformation behavior.

\begin{figure}[htbp]
    \centering
    \includegraphics[width=\linewidth]{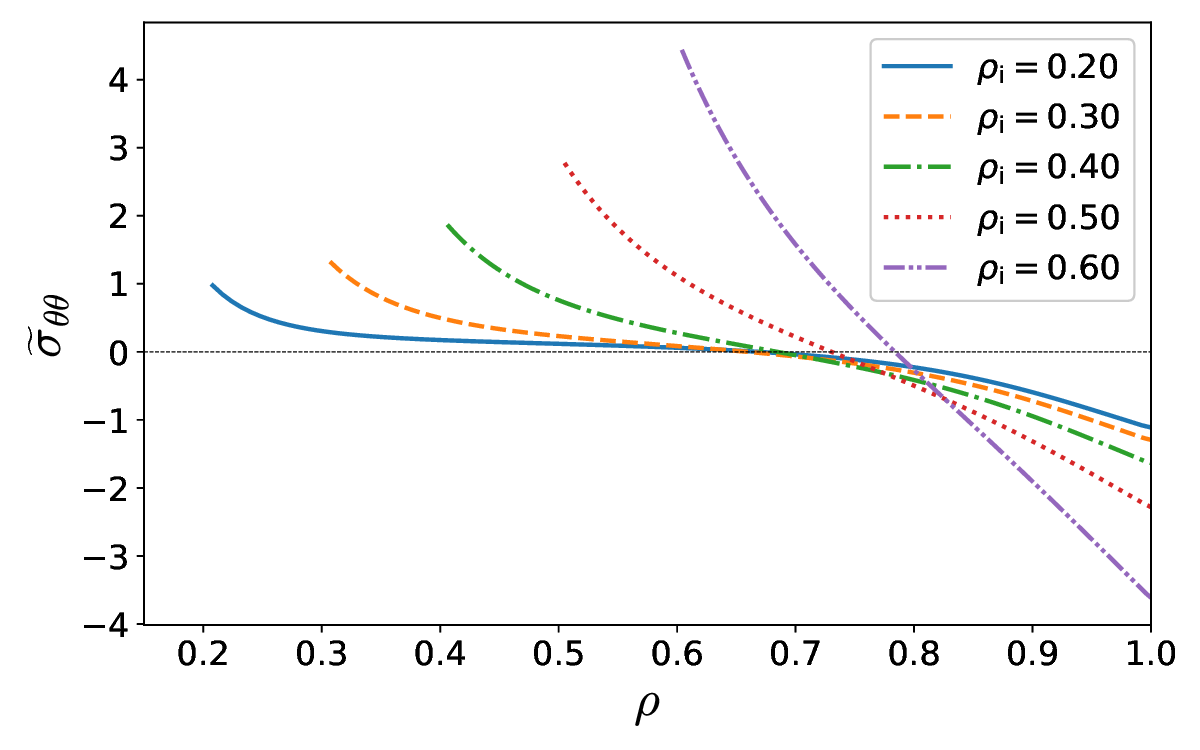}
    \includegraphics[width=\linewidth]{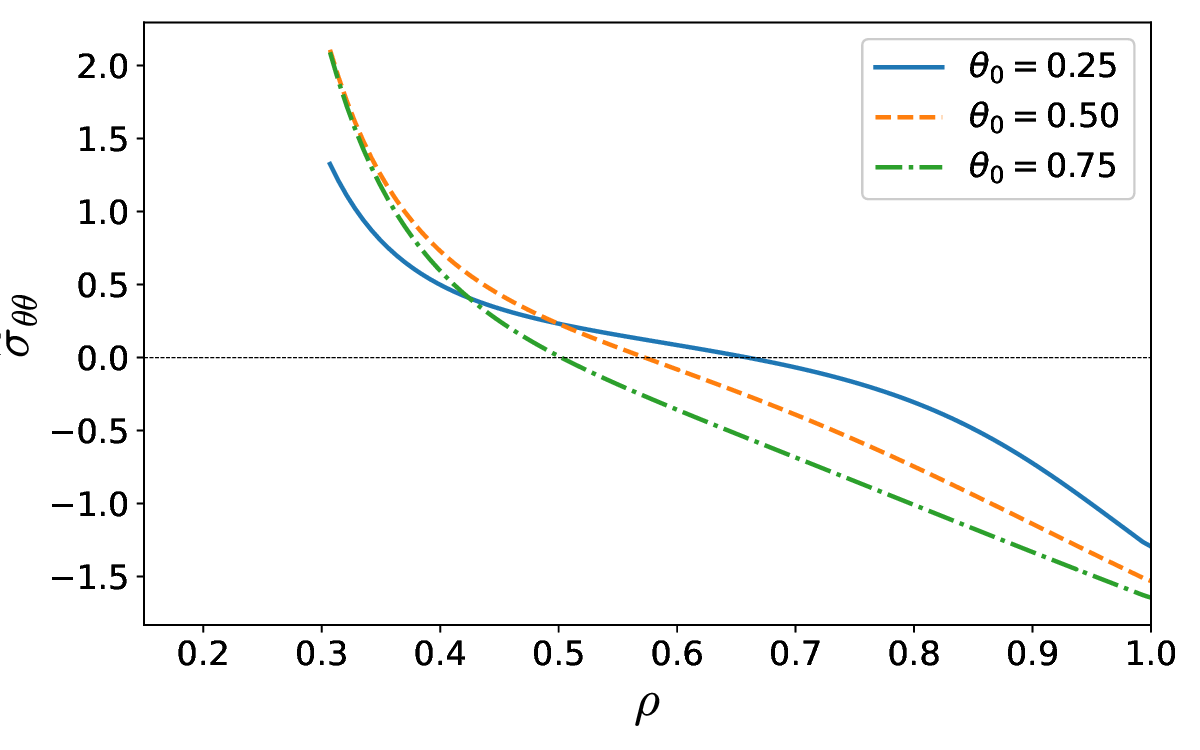}
    \caption{Plots of the (dimensionless) tensile stress along the loading line against $\rho$ for (top) various inner radii $\rhoi$ with $\theta_0=0.25$ and (bottom) various angles $\theta_0$ with $\rhoi=0.3$, where Poisson's ratio is fixed as $\nu=0.3$.}
    \label{fig:load_line}
\end{figure}
Now, let us check the profile of the tensile on the loading axis for various $\rhoi$ and $\theta_0$ in Fig.~\ref{fig:load_line}.
It is interesting that the tensile stress is not constant on this line, which is contrast to a solid disk~\cite{Timoshenko}.
These behaviors are also similar to the case for the concentrated loading cases~\cite{Okamura25}.

\section{Conclusion and discussion}
In this paper, we have successfully derived the displacement and stress of a two-dimensional elastic hollow disk subjected to a distributed diametric loading within the framework of elastodynamic theory.
Our results provide a detailed understanding of how the distribution of loading influences the locations of displacement and stress maxima and minima in the hollow disk.
By analyzing the loading region, we have clarified the dominant locations of tensile and compressive stresses, as well as how their distributions differ from the case of a concentrated load.

While our present study focuses on the static response, the framework can be extended to dynamic cases.
By carefully handling the inverse Laplace transform of the deformation and stress, as demonstrated in previous two-dimensional analyses~\cite{Jingu85, Sato24}, we can obtain their time evolution.
Although such calculations may be more involved, they represent an important next step in our research.

\section*{Acknowledgment}
The authors thank Shintaro Hokada for his comments and helps on visualization.
This work is partially supported by the Grant-in-Aid of MEXT for Scientific Research (Grant No.~24K06974 and No.~24K07193).


\end{document}